\RequirePackage{mathtools}

\documentclass[12pt]{iopart}
\usepackage{graphicx}
\usepackage{hyperref} 
\usepackage[numbers,sort&compress]{natbib}
\usepackage{caption}
\usepackage{subcaption}
\usepackage{braket}
\usepackage{booktabs}
\usepackage[table]{xcolor}

\usepackage[separate-uncertainty=true,exponent-product = \cdot]{siunitx}

\usepackage{csquotes}

\newcommand{\ybyso}[0]{$^{171}$Yb$^{3+}$:Y$_2$SiO$_5$}

\newcommand{\yso}[0]{Y$_2$SiO$_5$}
\newcommand{\pryso}[0]{Pr$^{3+}$:Y$_2$SiO$_5$}
\newcommand{\euyso}[0]{$^{151}$Eu$^{3+}$:Y$_2$SiO$_5$}
\newcommand{\eu}[0]{Eu$^{3+}$}
\newcommand{\pr}[0]{Pr$^{3+}$}

\newcommand{\ybiso}[0]{$^{171}$Yb$^{3+}$}

\newcommand{\ybtrans}{$^2$F$_{7/2}(0) \leftrightarrow ^2$F$_{5/2}(0)$}
\newcommand{\gstate}{$^2$F$_{7/2}(0)$}
\newcommand{\estate}{$^2$F$_{5/2}(0)$}

\DeclareMathOperator{\sgn}{sgn}
\DeclareMathOperator{\sinc}{sinc}
\DeclareMathOperator{\rect}{rect}
\DeclarePairedDelimiter\abs{\lvert}{\rvert}
\newcommand\dd[1]{\ \text{d}{#1}}

\begin{document}

\title{Broadband and long-duration optical memory in \ybyso{}}

\author{T. Sanchez Mejia$^1$, L. Nicolas,$^{1,2}$ A. Gelmini Rodriguez,$^1$ P. Goldner,$^3$ and M. Afzelius$^1$}

\address{$^1$Department of Applied Physics, University of Geneva, 1205 Geneva, Switzerland.

$^2$Department of Applied Physics, Aalto University, 02150 Espoo, Finland.

$^3$Chimie ParisTech, PSL University, CNRS, Institut de Recherche de Chimie Paris, Paris, France.  }
\ead{mikael.afzelius@unige.ch}

\vspace{10pt}

\begin{abstract}
Optical quantum memories are essential components for realizing the full potential of quantum networks. Among these, rare-earth-doped crystal memories stand out due to their large multimode storage capabilities. To maximize the multimode capacity in the time domain, it is key to simultaneously achieve large memory bandwidth and long optical storage time. Here, we demonstrate an atomic frequency comb optical memory in \ybyso{}, with a memory bandwidth of 250~MHz and a storage time of up to 125~$\mu$s. The efficiency reaches 20\% at short storage times, and 5\% at 125~$\mu$s. These results were enabled by an optimized optical pumping scheme, guided by numerical modelling. Our approach is specifically designed for future spin-wave storage experiments, with the theoretical bandwidth limit set at 288 MHz by the hyperfine structure of \ybyso{}. Additionally, we introduce an efficient method for synthesizing the optical pumping waveforms required for generating combs with tens of thousands of teeth, as well as a simple yet frequency-agile laser setup for optical pumping across a 10 GHz bandwidth.
\end{abstract}

%
%
%
%
%

\section{Introduction}\label{sec:intro}

Optical quantum memories using rare-earth (RE) ion doped crystals are one of the main candidates for building long-distance quantum networks \cite{Kimble2008,Wehner2018}, owing to their long optical and spin coherence times, and their ability to store many modes in a single quantum memory, see Ref. \cite{Tittel2025} and references therein. In this context, RE-doped crystals have achieved storage of single photons \cite{Clausen2012,Zhou2012,Seri2018,Seri2019}, entanglement \cite{Clausen2011,Saglamyurek2011,Ferguson2016,Rakonjac2021}, and have been used to generate entanglement between two quantum memories \cite{Usmani2012,lago2021,Liu2021} (for a more complete list of relevant work see Ref. \cite{Tittel2025}).

The temporal multimode capacity of an RE quantum memory generally exploits the large number of spectral channels in the absorption spectrum \cite{Tittel2010b}, i.e. the ratio of the optical inhomogeneous broadening to the homogeneous broadening (the inverse of the optical coherence time), which lets one store the phase and amplitude information of an incoming train of photons. Storage of many temporal modes thus requires large memory bandwidth and long optical coherence time (narrow homogeneous linewidth), to maximize this ratio. 

Many recent RE quantum memory demonstrations have employed the atomic frequency comb (AFC) scheme \cite{Afzelius2009a}, where a periodic comb of narrow absorption lines are created over the memory bandwidth $\Gamma_{\mathrm{AFC}}$, through optical pumping. The multimode capacity is proportional to the number of teeth in the AFC \cite{Afzelius2009a,Ortu2022}, where the width of each tooth is limited by the optical homogeneous linewidth. This concretizes the general argument above of how the multimode capacity depends on memory bandwidth and the optical coherence time.

Several parameters limit the bandwidth of a RE quantum memory, among those is the splitting between the hyperfine/Zeeman ground states that are used for optical pumping. Therefore, RE ions with large ground-state splittings can achieve larger bandwidths, which is the case for Kramers ions having large electronic spin. In Er and Nd systems, large memory bandwidths in the range of 100 MHz to 6 GHz have been demonstrated \cite{Tiranov2015a,TangZhouWangEtAl2015,Askarani2019,Craiciu2019,Liu2021}. However, the effective optical coherence times of these memories have been poor, typically in the 10 ns to 1 $\mu$s range.

To achieve both broadband and long duration optical storage, the Kramers ion \ybiso{} is a highly interesting alternative. Its hyperfine splitting is of the order of GHz \cite{Tiranov2018a,Kindem2018}, allowing large bandwidth in principle, while simultaneously having long optical coherence times thanks to a Zero First-Order Zeeman (ZEFOZ) effect at zero magnetic field \cite{Ortu2018,Kindem2020,Welinski2020,Nicolas2023}. Exploiting these unique features, we recently demonstrated up to $\SI{25}{\micro\second}$ optical storage of non-classical light with a memory bandwidth of $\SI{100}{\mega\hertz}$ \cite{Businger2020}.

To reach longer storage times, one can convert the optical coherence into a spin-coherence, using the spin-wave AFC scheme \cite{Afzelius2009a,Afzelius2010}. Demonstrations of spin-wave AFC quantum memories have so far relied on non-Kramers ions \pr{} (e.g. \cite{Ferguson2016,Rakonjac2021}) and \eu{} (e.g. \cite{Ma2021,Ortu2022b}), with bandwidths limited to $<10$ MHz due to their small nuclear hyperfine splits. A \ybiso{} spin-wave AFC memory could potentially reach over 100 MHz bandwidth, the largest bandwidth being the 10-MHz spin-wave demonstration in \ybyso{} \cite{Businger2020}, although several challenges remain in order to reach that goal.

One challenge is to devise an optical pumping scheme that allows broadband pumping in an inhomogeneously broadened system, taking into account the overlapping line spectrum due to the hyperfine splittings in the ground and excited states. With RE crystals, these schemes are necessary in order to assure that we excite ions on very specific transitions within the optical inhomogeneous broadening \cite{Nilsson2004,Rippe2005,Lauritzen2012,Chen2025b}, which are often referred to as class selection or class cleaning (CC) techniques. The effective bandwidth of the quantum memory is in practice limited by the frequency range over which these CC techniques work. For example, in \euyso{}, the frequently used CC scheme limits the bandwidth to $\SI{5.7}{\mega\hertz}$ \cite{Ortu2022}, despite that the nuclear hyperfine splitting ranges from 35 MHz to 101 MHz.

In this work, we propose a new CC scheme for \ybyso{} that would in principle allow us to achieve an AFC spin-wave memory over a 288 MHz bandwidth. This scheme would use the 3025.5 MHz hyperfine transition for long-duration storage, with expected spin coherence times in the range of 1 - 10 ms at the zero-field ZEFOZ point and a temperature of around 3 K \cite{Ortu2018,Nicolas2023}. This high-frequency spin transition can be efficiently manipulated using a lumped-element microwave resonator \cite{Nicolas2023}, a key requirement for long duration storage. The CC scheme is simulated using a simple optical pumping simulator \cite{Chen2025b}, showing that the bandwidth is limited only by an excited state hyperfine split. Experimentally, we demonstrate that the optical pumping scheme can spin polarise $\SI{80}{\percent}$ of the total population into a single hyperfine state, thereby increasing the optical depth of the relevant AFC transition by a factor of 3.3. Exploiting the increased optical depth, the short-delay AFC storage efficiency reaches around 20\%, which is constant throughout the 288 MHz bandwidth. Leveraging the long optical coherence time in \ybyso{}, we demonstrate AFC echoes for up to 125 $\mu$s delays, with a 5\% efficiency at 125 $\mu$s, which corresponds to an effective AFC coherence time of $T_2^{\mathrm{AFC}} = \SI{348 \pm 15}{\micro\second}$s, the longest achieved in any RE crystal to date. 

These results can be attributed to several developments with respect to our previous AFC storage experiments in \ybyso{} \cite{Businger2020,Businger2022}. The \ybiso{} doping concentration has been reduced from 5 ppm to 2 ppm, resulting in a longer optical coherence time of $1050~\mu\mathrm{s}$ at around 3 K \cite{Nicolas2023,Chiossi2024}. The lower concentration also reduces the spin-spin flip-flop relaxation rate, making the optical pumping of the hyperfine states more efficient. A key technical development is a frequency-agile optical setup for coherent phase and amplitude control over a bandwidth of 10 GHz, using a single frequency-locked laser and a single electro-optic modulator (EOM). We also developed a new, faster method for synthesizing the AFC preparation pulse, which allows for a finer parameter estimation in order to maximize the AFC echo efficiency, an important practical aspect when generating broadband AFCs consisting of tens of thousands of comb teeth.

\section{Single class selection and spin polarisation over 288 MHz in \texorpdfstring{\ybyso}{Yb:YSO}}

\subsection{Selecting the \texorpdfstring{$\Lambda$}{Lambda}-system for AFC spin-wave storage}
\label{sec:lambda_system}

The storage of an optical pulse as a spin excitation relies on the transfer of the optical coherence into a spin coherence, using an optical control field. The input field and the control field both couple to a common excited state and two hyperfine ground states, in a so-called $\Lambda$ configuration. In AFC spin-wave memories, the transfer is done with a coherent population inversion pulse, typically an adiabatic pulse for an efficient and uniform transfer over the entire AFC bandwidth \cite{Rippe2005,Minar2010,Ortu2022b,Chen2025}. In practice, the bandwidth of the AFC spin-wave memory is limited both by the optical Rabi frequency on the control pulse transition \cite{Chen2025}, and the bandwidth of the class cleaning sequence \cite{Chen2025b}. Furthermore, the input-field transition, in our case the transition on which the AFC is created, should also be as strong as possible, to increase the optical depth.

The \ybyso{} system presents two advantages with respect to other RE ions in this context. First, the optical dipole moment of the relevant optical transition \ybtrans{} is one of the strongest in \yso{} \cite{Welinski2016}. Second, the relative transition strengths, when coupling different hyperfine levels in the ground and excited states (the branching ratio matrix elements), are polarisation dependent \cite{Nicolas2023}. In particular, the optical dipole moments for light polarised along the two dielectric axes $D_2$ and $b$ are similar in \ybyso{} \cite{Welinski2016}, but the branching ratio tables have their strongest elements along either the anti-diagonal or the diagonal directions of the matrix, as shown in Table \ref{tab:branching-tables}. This allows a larger freedom when selecting a $\Lambda$-system for quantum storage and other similar schemes, and the optical dipole can be maximized for both transitions simultaneously. In this work, we consider a $\Lambda$ system where the input and control transitions use the strong transitions $\ket{4_g} - \ket{1_e}$ and $\ket{1_g} - \ket{1_e}$, polarised along the $D_2$ and $b$ axes, respectively, see Table \ref{tab:branching-tables} and Fig. \ref{fig:CC_scheme}(a). The corresponding spin transition $\ket{1_g} - \ket{4_g}$ has a frequency of 3025.5~MHz, which one can efficiently manipulate using a lumped-element MW resonator, even for long crystals of length 10-15 mm \cite{Nicolas2023}. In Ref. \cite{Businger2020}, an optical Rabi frequency of 2~MHz was obtained on the $\ket{4_g} - \ket{1_e}$ transition for a peak power of $\SI{300}{\milli\watt}$. A similar Rabi frequency can be expected on the $\ket{1_g} - \ket{1_e}$ transition, which would yield around 90\% efficiency over 100~MHz bandwidth for the adiabatic control pulse considered in the analysis of Ref. \cite{Businger2020}. However, a more refined analysis of the optical and spin manipulation should be made in a future work, using a model like the one presented in Ref. \cite{Chen2025}, which would allow one to accurately model the effective memory bandwidth under different experimental conditions in \ybyso{}.

\begin{table}
\begin{center}
\begin{tabular}{c|cccc||c|cccc}
 $\mathbf{E}||D_2$ & $\mathbf{\ket{1_e}}$ & $\mathbf{\ket{2_e}}$ & $\mathbf{\ket{3_e}}$ & $\mathbf{\ket{4_e}}$ & $\mathbf{E}||b$ & $\mathbf{\ket{1_e}}$ & $\mathbf{\ket{2_e}}$ & $\mathbf{\ket{3_e}}$ & $\mathbf{\ket{4_e}}$ \\
\midrule
$\mathbf{\ket{1_g}}$ & 0.15 & 0.06 & 0.08 & 0.71 & $\mathbf{\ket{1_g}}$ & \cellcolor{gray!30}0.54 & 0.19 & 0.03 & 0.24 \\
$\mathbf{\ket{2_g}}$ & 0.06 & 0.19 & \cellcolor{gray!30}0.71 & 0.04 & $\mathbf{\ket{2_g}}$ & 0.21 & 0.57 & 0.21 & 0.01 \\ 
$\mathbf{\ket{3_g}}$ & 0.07 & 0.71 & 0.16 & \cellcolor{gray!30}0.06 & $\mathbf{\ket{3_g}}$ & 0.01 & 0.18 & 0.66 & 0.15 \\
$\mathbf{\ket{4_g}}$ & \cellcolor{gray!30}0.72 & 0.04 & 0.05 & 0.19 & $\mathbf{\ket{4_g}}$ & 0.23 & 0.06 & 0.10 & 0.61 \\
\end{tabular}
\caption{Relative transition strengths for different optical-hypefine transitions between the ground \gstate{} and excited \estate{} states in \ybyso{}, for the light field polarised along dielectric axes $D_2$ and $b$ in \yso{}. The $\Lambda$-system and CC frequencies proposed in this work are highlighted by the shaded boxes, see Secs \ref{sec:lambda_system} and \ref{sec:pumping_simulation} for details. The matrices are reproduced from the Supplementary Information of Ref. \cite{Nicolas2023}. See Fig. \ref{fig:CC_scheme} for the order and labelling of the hyperfine states.}
\label{tab:branching-tables}
\end{center}
\end{table}

\subsection{Simulation of the broadband class cleaning scheme}
\label{sec:pumping_simulation}

\begin{figure}
    \centering
    \includegraphics[width=1\linewidth]{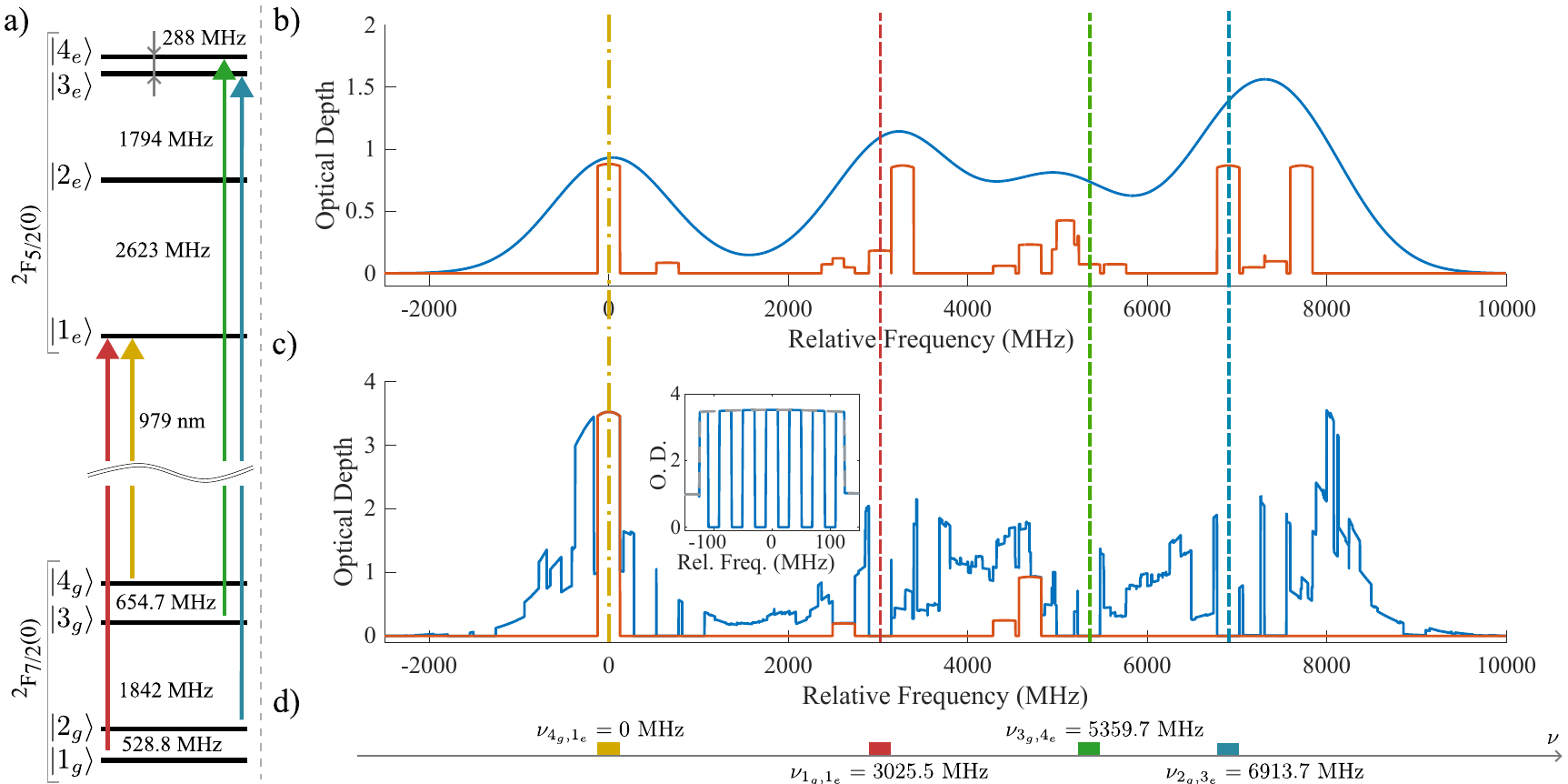}
    \caption{a) Energy level scheme for site 2 of \ybyso{} \cite{Welinski2016,Tiranov2018a}, with the four proposed CC transitions. b) Numerical simulation of the un-pumped inhomogeneous absorption spectrum shown in blue. The contribution to the absorption from the frequency class of the CC scheme is shown in red. c) Numerical simulation of the absorption spectra after CC and SP steps, over a bandwidth of 250 MHz, using the same colour coding as in b). Note that for the selected frequency class (red curve), we only see four distinct transitions, consistent with the optical pumping of its hyperfine population into the $\ket{4_g}$ state. In d), we show the CC frequencies which are scanned over 250 MHz around their central transitions, see text for details. Inset: an atomic frequency comb prepared in the $\nu_{4_g,1_e}$  anti-hole after the CC and SP steps.}
    \label{fig:CC_scheme}
\end{figure}

The CC procedure is a method to ensure that we excite specific transitions, despite that the inhomogeneous broadening causes overlap of different optical-hyperfine transitions in the spectrum. The general idea is to pump away ions that absorb light on unwanted transitions \cite{Nilsson2004,Rippe2005,Lauritzen2012,Chen2025b}. Specifically, in our case, we need to ensure that the two frequencies employed in the $\Lambda$-system only excite the $\ket{4_g} - \ket{1_e}$ and $\ket{1_g} - \ket{1_e}$ transitions. The isolated low-frequency peak in the spectrum, see Fig. \ref{fig:CC_scheme}(b), absorbs mostly on the $\ket{4_g} - \ket{1_e}$ transition, i.e. mostly a single frequency class contributes to this absorption peak. This is the main reason we select this peak as one of the transitions in the $\Lambda$-system. However, the second transition $\ket{1_g} - \ket{1_e}$ sits in a denser part of the spectrum, and other ions within the inhomogeneous broadening absorb at the same frequency, but on other transitions, these are defined as other frequency classes \cite{Chen2025b}.

A simple CC method consists in choosing a set of transitions that excite all the ground states of the selected frequency class \cite{Chen2025b}. Other frequency classes will not resonate with all these frequencies and their population are pumped away to hyperfine states that are not absorbing at these particular frequencies, provided that optical pumping is efficient. We thus need to select two additional frequencies that excite the $\ket{2_g}$ and $\ket{3_g}$ hyperfine states, to complete the CC scheme. In this work, we consider the transitions $\ket{2_g} - \ket{3_e}$ and $\ket{3_g} - \ket{4_e}$, as shown in Fig. \ref{fig:CC_scheme}(a).  The frequencies of the CC transitions are then $\nu_{4_g,1_e} = 0~\mathrm{MHz}$, $\nu_{1_g,1_e} = 3025.5~\mathrm{MHz}$, $\nu_{3_g,4_e} =5359.7~\mathrm{MHz}$ and $\nu_{2_g,3_e} = 6913.7~\mathrm{MHz}$, relative to the $\ket{4_g} - \ket{1_e}$ input transition in the $\Lambda$-system. We also assume that the $\nu_{4_g,1_e}$ frequency is centred on the isolated low-frequency absorption peak in the spectrum, see Fig. \ref{fig:CC_scheme}(b), at the absolute frequency of 306264.0~GHz (978.8694 nm in vacuum).

Following the CC step, one can spin polarise all the population into a given hyperfine state. In our case, we spin polarise into the $\ket{4_g}$ state, in order to maximize the optical depth on the $\ket{4_g} - \ket{1_e}$ transition, which is used for absorbing the input light in the AFC storage scheme. The CC and SP preparation steps can be simulated with the simple numerical model presented in Ref. \cite{Chen2025b}. In Fig. \ref{fig:CC_scheme}(b)-(c), we show the resulting inhomogeneous absorption spectrum before and after applying the CC and SP steps using the proposed CC frequencies. The blue curves show the total spectrum with contributions from all ions in the inhomogeneous broadening, while the red curves show the contribution to the absorption spectrum from the selected frequency class. All CC and SP frequencies are scanned over 250 MHz around the central frequencies given above. As seen in Fig. \ref{fig:CC_scheme}(c), the CC and SP steps create a flat and highly absorbing anti-hole with a width of 250 MHz at the $\nu_{4_g,1_e}$ frequency, where only the selected frequency class absorbing on the $\ket{4_g} - \ket{1_e}$ transition contributes to the absorption. The simulated spectrum also shows 250-MHz wide transmission windows centred on the $\nu_{1_g,1_e}$, $\nu_{3_g,4_e}$ and $\nu_{2_g,3_e}$ frequencies. Note that in an AFC spin-wave storage experiment, the optical control field would be applied in the transparency window around $\nu_{1_g,1_e}$.

The CC step can only be achieved over a certain maximum bandwidth, beyond which the other frequency classes cannot be entirely pumped away. By simulating successively larger bandwidths, it is possible to deduce the maximum bandwidth by studying the resulting absorption spectrum, as discussed in Ref. \cite{Chen2025b}. Using the proposed CC frequencies, we deduce that the maximum bandwidth is 288 MHz, limited by the $\ket{3_e} - \ket{4_e}$ excited state hyperfine split, as discussed in Sec. \ref{sec:BWlimit}. In comparison, our simulations of the CC scheme used in Ref. \cite{Businger2022}, shows a maximum bandwidth of 125 MHz. That scheme is based on the same $\Lambda$-system as in this work, but employs two different additional CC frequencies, which reduces the bandwidth. We also note that in Ref. \cite{Businger2020}, we considered yet another set of CC frequencies, which also achieves a maximum bandwidth of 288~MHz. However, that particular CC system involves another $\Lambda$-system, where the control field excites a transition with significantly lower optical dipole moment. That $\Lambda$-system limits the Rabi frequency of the optical control field, which in turn, reduces the achievable AFC memory bandwidth. We believe that the $\Lambda$-system and the set of CC frequencies proposed in this work represents the most promising scheme for achieving a broadband AFC spin-wave memory in \ybyso{} in theory.

\section{Efficient numerical synthesis of the AFC preparation pulse} 
\label{sec:AFC_numerical_signal_tricks}

The optimal AFC shape, for a finite optical depth, is a comb with square-shaped teeth \cite{Bonarota2010}. The efficiency of the AFC echo for such a comb is given by

\begin{equation}
    \eta_{\mathrm{AFC}} = \tilde{d}^2 e^{-\tilde{d}} \sinc^2(\pi / F),
    \label{eq:afc_echo_efficiency}
\end{equation}

\noindent where $\tilde{d} = d / F$ is the mean optical depth, given a peak optical depth $d$, and $F$ is the finesse of the AFC \cite{Afzelius2009a}. In Ref. \cite{Bonarota2010}, it was shown that, for a fixed peak optical depth $d$, there is an optimal finesse $F_{\mathrm{opt}} = \pi/\arctan(2\pi/d)$ that maximizes the AFC echo efficiency. We refer to this as the optimum comb and the goal of any AFC preparation method should be to attain this theoretical optimum, for a given finite $d$.

The creation of an AFC involves optical pumping at different frequencies in the inhomogeneously broadened transition, i.e. spectral hole burning. To burn the optimum square-shaped AFC, the power spectral density (PSD) of the AFC preparation pulse should also be a square-shaped optical frequency comb. It should have perfectly rectangular bands of power in between the AFC teeth and ideally no optical power anywhere else, to avoid burning away any optical depth on the teeth. The edges of the square teeth of the PSD should be as sharp as possible, so that the only limiting factor to the spectral resolution of the AFC is the homogeneous linewidth. The finesse of the optical comb should be set to $\tilde{F} = (1 - F^{-1})^{-1}$, in order to burn an AFC of finesse $F$. That is, if we denote $\hat{s}(f) = u(f) e^{i \theta(f)}$ the Fourier transform of the AFC-burning signal $s(t)$, the PSD $u^2(f) = \abs{\hat{s}(f)}^2$ of the ideal AFC preparation pulse should be

\begin{equation} \label{eq:ideal_AFC_burning_PSD}
    u^2(f) \propto \sum_{n = 0}^{N - 1} \rect \frac{f - \sigma /2 - n \Delta}{\sigma} ,   
\end{equation}

\noindent where $\Delta$ is the comb periodicity, $N = \Gamma_{\mathrm{AFC}} / \Delta$ the number of teeth, $\sigma = \Delta /\tilde{F}$ the width of the preparation pulse's teeth, and the rectangular function $\rect(x) = 1$ for $\abs{x} < 1/2$ and $0$ otherwise.

A straightforward way to achieve this is to optically pump in between each tooth of the comb sequentially, see e.g. Ref \cite{Clausen2011}, i.e. one tooth at a time with $N$ subsequent frequency chirped pulses $s_n(t)$ (i.e. chirps) of duration $T$ that are shifted in frequency by $\Delta$ from each other. In practice, this can be achieved by sending the chirps as a sequence of radio-frequency (RF) signals to an acousto- (AOM) or electro-optic (EOM) modulator -- see Sec. \ref{sec:optical_pumping_signal_experimental_setup} for an explanation of our experimental implementation. However, when working in a regime where the AFC is broadband and the storage time is long, that is when the number of teeth $N$ is large, this is not feasible, as the duration of the burning sequence $N T$ would eventually exceed the spin population relaxation time $T_1$. This becomes especially problematic at longer storage times, as the duration $T$ needs to be larger than $1 / \Delta$ to allow for enough spectral resolution.

To overcome this problem, a method was proposed in Ref. \cite{Jobez2016} whereby all the teeth are burnt simultaneously by a single pulse. This is done by summing up all of the individual chirps $s_n(t)$ in the time domain, such that one AFC burning cycle only takes a time $T$ instead of a time $N T$. This results in an analytical formula for the waveform $s(t)$ that is fast to compute numerically. However, with this method, since all of the chirps are summed up in phase, the resulting signal will resemble a Dirac comb, with most of the energy concentrated at rephasing times that are multiples of $1 / \Delta$, such that its peak temporal amplitude scales linearly with the number of teeth. When using this method with an AOM or EOM, this peak amplitude is limited by the maximum modulation voltage of the device, such that the average power in the pulse effectively scales inversely with the number of teeth $1/N$ \cite{Businger2022}.

An alternative method, which we refer to here as the "circular permutation" method, was proposed in Ref. \cite{Businger2022}, that eliminates both the $\mathcal{O}(N)$ scaling of the burning duration and the $\mathcal{O}(N)$ scaling of the temporal signal amplitude. As described in their Supplementary Note I, their method is based on applying an appropriate temporal delay on each individual chirp $s_n(t)$, to prevent the constructive interference effect when summing them. However, this approach is numerically intensive, as the time it takes to compute the numerical waveform $s(t)$ scales linearly with the number of teeth, owing to the fact that the individual chirps have to be summed up numerically as no closed-form expression for the sum is known. In this work, we improve on this method by removing the $\mathcal{O}(N)$ scaling of the signal computation time, while also retaining the constant-$N$ amplitude and burning duration. 

Instead of working with chirps in the time domain, the idea is to work in the frequency domain directly and construct a spectral phase $\theta(f)$ which maximizes the root mean square (RMS) power of the signal at constant peak temporal amplitude, while setting the power spectral density to be exactly the frequency comb Eq. \ref{eq:ideal_AFC_burning_PSD} with the desired $\Delta$ and finesse. One approach to construct such a $\theta(f)$ analytically is to approximate the Fourier integral of the temporal signal using the stationary phase approximation \cite{Fowle1964}. As shown in the \ref{sec:AFC_spectral_phase}, what we obtain is the spectral phase 
\begin{equation} \label{eq:phase_new_method}
    \frac{1}{2\pi} \theta(f) = \left((f \backslash \Delta)\Delta \right)^2 \frac{1/\Delta}{2 \Gamma_{\mathrm{AFC}}} + \left(f \% \Delta\right)^2 \frac{T}{2\sigma},
\end{equation}
where $a \backslash b = \lfloor a / b\rfloor$ is the the integer division, and $a \% b$ is the remainder of $a / b$, with $\sigma$ the width of the teeth. The first term is constant over one $\Delta$ ($f \backslash \Delta$ indexes the teeth from $0$ to $N-1$), while the second term is periodic with period $\Delta$ ($f \% \Delta$ is a sawtooth wave that ramps from $0$ to $\Delta$). The total frequency-domain AFC-burning signal, whose PSD is given by Eq. \ref{eq:ideal_AFC_burning_PSD}, can therefore be written as
\begin{equation*} 
    \hat{s}(f) \propto \sum_{n = 0}^{N - 1} e^{i \pi \frac{n^2}{N}}  e^{i \pi \left(f - n \Delta\right)^2 \frac{T}{\sigma}} \rect \frac{f - \sigma / 2 - n \Delta}{\sigma}.   
\end{equation*}

\begin{figure}
    \centering
    \includegraphics[width=1\textwidth]{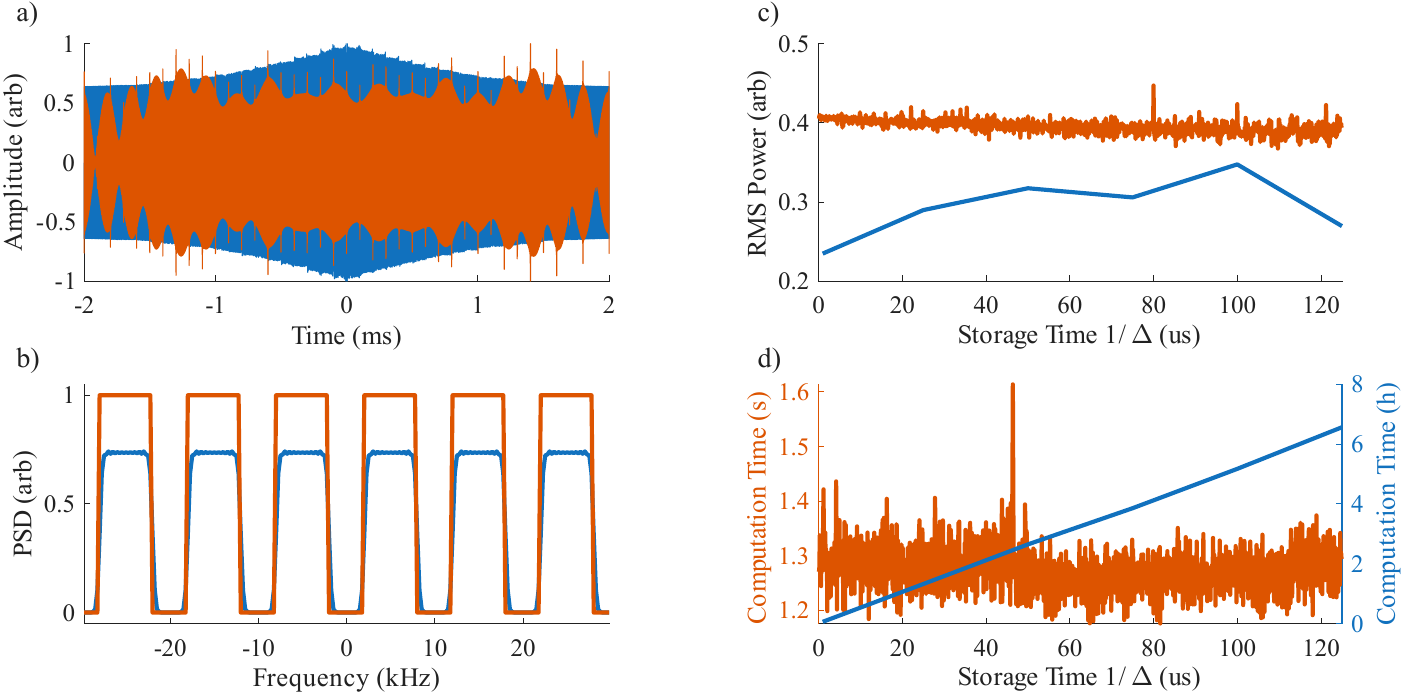}
    \caption{In this figure we compare two methods for synthesizing the AFC preparation pulse waveform, the circular permutation \cite{Businger2022} (blue) and the current frequency-domain method (red). a) Amplitude-normalized time-domain AFC preparation waveform for a storage time of $1/\Delta = \SI{100}{\micro\second}$ and a bandwidth of $\Gamma_{\text{AFC}} = \SI{250}{\mega\hertz}$. b) Power spectral density (PSD) of the signals in (a), zoomed on $\SI{60}{\kilo\hertz}$ out of the $\SI{250}{\mega\hertz}$ band. c) RMS power and d) numerical waveform computation time as a function of the programmed storage time.}
    \label{fig:AFC_preparation_waveform}
\end{figure}

The circular permutation method and this frequency-domain method are compared in Fig. \ref{fig:AFC_preparation_waveform}. The frequency-domain method allows for a slightly higher RMS power than the circular permutation method for a fixed temporal peak amplitude, as seen in Fig. \ref{fig:AFC_preparation_waveform} (b) and (c). The main benefit of the frequency-domain method is that the waveform is much faster to compute numerically (constant-$N$ time scaling vs linear scaling, resp, as seen in Fig. \ref{fig:AFC_preparation_waveform} (d)). In practice, most of the short constant-$N$ computation time is spent computing the inverse discrete Fourier transform of the signal's spectrum, which can be done quickly using the fast Fourier transform algorithm. Computing the power spectral density and the spectral phase takes a negligible amount of time, as they are respectively a simple quadratic function of $f$ and a simple rectangular mask. This is particularly useful when optimizing the finesse of the signal (as mentioned in Sec. \ref{sec:bb_eff_echoes}) at long storage times, which is something that would take days with the circular permutation method because of the time required to compute the waveforms for different finesses.

\section{Experimental implementation}\label{sec:experimental_setup}

\begin{figure}
    \centering
    \includegraphics[width=1\linewidth]{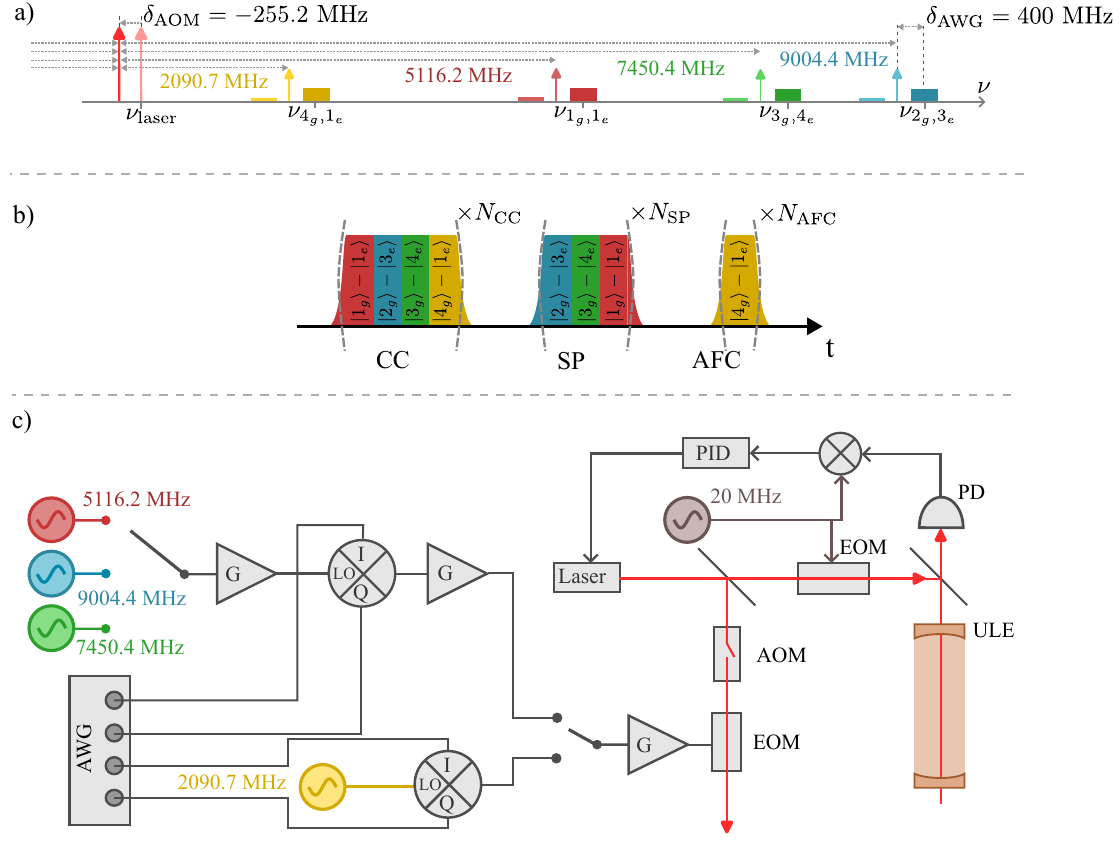}
    \caption{a) Frequency spectrum of the laser after phase modulation. b) Preparation sequence. CC: Class cleaning, SP: spin polarisation, AFC: AFC preparation. c) Schematic of the microwave and radio-frequency circuit used to drive the optical phase modulator (EOM) to shape the spectrum of the preparation laser. AWG: Arbitrary Waveform Generator, PID: Proportional–integral–derivative controller, AOM: acousto-optic modulator, G: Amplifier, (I,LO,Q): IQ-mixer, PD: photodiode.}
    \label{fig:exp_setup}
\end{figure}

\subsection{One laser and one phase modulator driven by complex MW pulses}\label{sec:optical_pumping_signal_experimental_setup}

The preparation sequence required for AFC storage experiments involves a complex sequence of CC, SP and AFC preparation pulses at different frequencies, see Fig. \ref{fig:CC_scheme}, with complete amplitude and phase control, as outlined in the previous sections. For non-Kramers ions with small nuclear hyperfine splittings, such as \pr{} and \eu{}, the required frequency shifts can easily be achieved with a single AOM and a single laser. The large hyperfine splittings in Kramers ions like \ybiso{} require another approach to generate the optical frequencies, which can span up to 10 GHz. Experimentally, the complexity is inevitable in such a setup, but one can choose where this complexity is located. For example, one can choose to use multiple lasers, which eases the requirements on the electronics and modulators but transfers the complexity to the frequency off-set stabilization of multiple lasers, e.g. \cite{Businger2020}. In this work, we use a single laser and a single EOM phase modulator, alleviating the complexity of the physical experimental setup by shifting the complexity to the generation of the microwave (MW) and RF signals. The general idea is to generate all optical frequencies by sideband modulation of the laser, using a combination of local oscillators (LOs) in the MW range to achieve large frequency shifts, an arbitrary waveform generator (AWG) for precise phase and amplitude control of each frequency, and IQ (in-phase and quadrature) mixers in the MW regime, see Fig. \ref{fig:exp_setup}(a)-(c), as described in detail below.

The laser is locked on a high-finesse optical cavity made of ultra-low expansion (ULE) glass, for long-term stability and line narrowing, see Fig. \ref{fig:exp_setup}(c). It is locked on a mode that has its resonance 2235.5~MHz below the low-frequency absorption peak in the spectrum, which avoids unwanted optical pumping at the laser frequency. To generate the required frequencies $\nu_{4_g,1_e}$, $\nu_{1_g,1_e}$, $\nu_{3_g,4_e}$ and $\nu_{2_g,3_e}$, we use four different continuous-wave LOs in the MW regime, tuned to the detuning between the laser and the relevant transition. These generate sidebands at the corresponding frequency shifts in the EOM, of which the high-frequency sideband is used for optical pumping, while the low-frequency sideband is far-detuned from the optical spectrum. Since each frequency has its own tuneable LO, any transition can be reached across the entirety of the absorption spectrum, spanning 10 GHz in total.

Arbitrary phase and amplitude control of each frequency is achieved using an IQ mixer, see Fig. \ref{fig:exp_setup}(c). An AWG outputs the in and out of phase quadrature signals of the desired waveform at the carrier frequency $\delta_{\mathrm{AWG}}$, which is fed into the I and Q ports of the IQ mixer, together with the MW LO signal. The IQ mixer is tuned to maximize the high-frequency $+\delta_{\mathrm{AWG}}$ sideband, while suppressing the leakage of the LO and the low-frequency sideband. The high-frequency sideband is tuned to the desired frequency, $\nu_{4_g,1_e}$, $\nu_{1_g,1_e}$, $\nu_{3_g,4_e}$ or $\nu_{2_g,3_e}$. For the CC/SP pulses, the AWG outputs a waveform with rectangular amplitude envelope and a linear frequency chirp. For the AFC preparation pulses, the AWG outputs a waveform with a comb-shaped power spectral density, as described in Sec. \ref{sec:AFC_numerical_signal_tricks}. 

The different frequency signals are combined using switches, amplified and sent to the EOM. A single-pole, quadruple-throw broadband MW switch combines the signals from the three LO sources corresponding to the $\nu_{1_g,1_e}$, $\nu_{2_g,3_e}$, $\nu_{3_g,4_e}$ frequencies. In between the CC/SP pulses, the switch is set to the unused port. A single-pole, double-throw switch is used to alternate between the signals coming from the two IQ mixers in the CC/SP and AFC branches. Having separate IQ mixers for the CC/SP and AFC signal generation allows for a better optimization of the IQ mixers, in particular for the AFC pulses with their complex phase and amplitude modulation. A single-pass AOM is used for slow amplitude modulation of the signal, to prevent light from reaching the crystal when input light is to be sent to the crystal for storage. The frequency shifts due to the AWG modulation, $\delta_{\mathrm{AWG}}$, and the AOM, $\delta_{\mathrm{AOM}}$, must be accounted for when tuning the LO frequencies, in order to reach the desired frequencies. The specific frequencies used in this work are shown in Fig. \ref{fig:exp_setup}(a).

\subsection{Memory setup}
\label{sec:optical_crystal}

The \ybyso{} memory crystal was grown using the Czochralski method, with a \ybiso{} doping concentration of $2\pm0.5$ ppm (isotopic purity of 95 \%). The crystal has dimensions 3.1 mm x 3.9 mm x 15.8 mm along the $b$, $D_2$, and $D_1$ dielectric axes, respectively. It is cooled to a temperature of $3.1~\mathrm{K}$ in a pulse tube cooler. All the optical beams propagate along the crystal $D_1$ axis, with their linear electric-field polarisation in the $D_2 - b$ plane, see discussion in Sec. \ref{sec:lambda_system}. The relatively long propagation axis compensates for the low absorption coefficient, due to the low doping concentration. The absorption coefficient is $\alpha = 0.158/\mathrm{cm}$ at the $\nu_{4_g,1_e}$ frequency in the spectrum, corresponding to an optical depth of $\alpha L = 0.25$ for the length $L = 15.8~\mathrm{mm}$. To further increase the optical depth, we use a quadruple-pass configuration \cite{Businger2020}, resulting in an effective optical depth at $\nu_{4_g,1_e}$ of 0.97, just slightly short of the expected $4\alpha L = 1$, as seen in Fig. \ref{fig:rep}(a).

The experiments employed two beams, one with a wider beam waist for the SP pulses (preparation mode), and another with a smaller beam waist for measuring the absorption spectrum and for the AFC storage experiments (input mode). The input mode beam waist radius was 60 $\mu$m. The preparation mode radius was 120 $\mu$s, to achieve a homogeneous optical pumping across the input mode within the crystal.

For the polarisation of the preparation mode in the crystal, we can look at the relative transition strengths of Table \ref{tab:branching-tables}. Transition $\ket{2_g}-\ket{3_e}$ is strongest in the $D_2$ axis, with transition $\ket{1_g}-\ket{1_e}$ strongest in the $b$ axis while transition $\ket{3_g}-\ket{4_e}$ is weak in both axes. After some tests, we found that a linearly polarised light at 45$^{\circ}$ in the $b - D_2$ plane offered a compromise in transition strength. The input pulse, and the laser frequency scans used to make Figs. \ref{fig:rep}, \ref{fig:bdw}, and \ref{fig:teeth}, use light polarised in the $D_2$ axis.

Because the coherence times of our crystal are detrimentally affected by vibrations, we were already using anti-vibration mounts in the previous experiment \cite{Businger2022}. This anti-vibration mount is made of two copper feet linked by a loose bundle of very thin copper strands. This allows for the dampening of vibrations while slightly sacrificing the thermal conductivity of the mount. To further improve the performances of our quantum memory, we are now using the vibrations of our pulse-tube cryostat to directly trigger our experiment. Using a well-placed ceramic piezoelectric stack, we can trigger our delay generator to start the preparation of the AFC at the moment in the cryostat cycle with the least vibrations.

\section{Experimental results}
\subsection{Efficient spin polarisation to increase optical depth}
\begin{figure}
    \centering
    \includegraphics[width=1\linewidth]{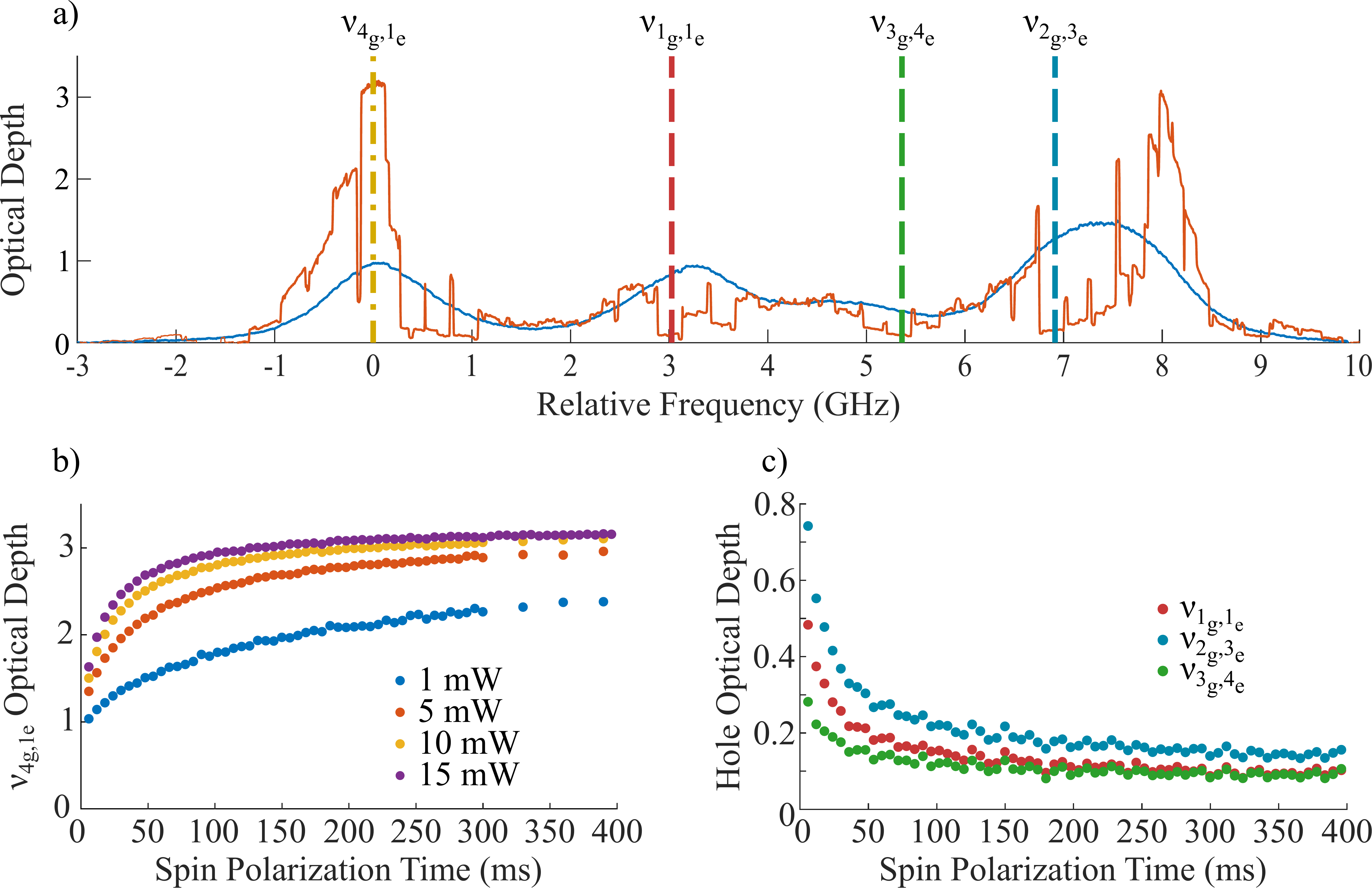}
    \caption{a) Experimental absorption spectrum before (blue curve) and after (red curve) the SP, for a bandwidth of 250 MHz. b)  The optical depth of the 250 MHz-wide $\ket{4_g}-\ket{1_e}$ anti-hole as a function of SP duration, for different optical powers before the cryostat. c) Optical depth in the transmission holes created by the SP pulses, as a function of the SP duration, for an optical power of 15 mW and a bandwidth of 250 MHz.}
    \label{fig:rep}
\end{figure}
In a first set of experiments, we study the efficiency of the optical pumping of ions into the $\ket{4_g}$ state through the SP step. In these experiments, we did not include the CC step. For the AFC echo experiments presented later in the paper, the CC step is not required; it is only required for spin-wave storage, which was not studied experimentally in this work. We also recall that the isolated low-frequency absorption peak in the spectrum almost entirely absorbs on the $\ket{4_g}-\ket{1_e}$ transition, see Fig. \ref{fig:CC_scheme}(a), which is the transition used for the AFC echo experiments.

The SP pulse train consists of a set of three 2-ms long pulses, one pulse for each frequency $\nu_{1_g,1_e}$, $\nu_{2_g,3_e}$, $\nu_{3_g,4_e}$, which are repeated for the total SP duration. Figure \ref{fig:rep}(a) shows the resulting absorption spectrum after the SP, for an optical power of 16 mW before the cryostat and for a total SP duration of 300 ms (i.e 50 repetitions). We show an increase in the optical depth at frequency $\nu_{4_g,1_e}$ from the unpolarised 0.97 to 3.2. Figure \ref{fig:rep}(b) shows that, at high power, the optical depth of the $\ket{4_g}-\ket{1_e}$ transition starts to saturate around 300 ms. This duration is approximately the same as the measured fast spin population lifetime, see \ref{sec:T1}. Figure \ref{fig:rep}(c) shows the optical depth in the three transmission windows created through the SP.

\begin{figure}
    \centering
    \includegraphics[width=0.8\linewidth]{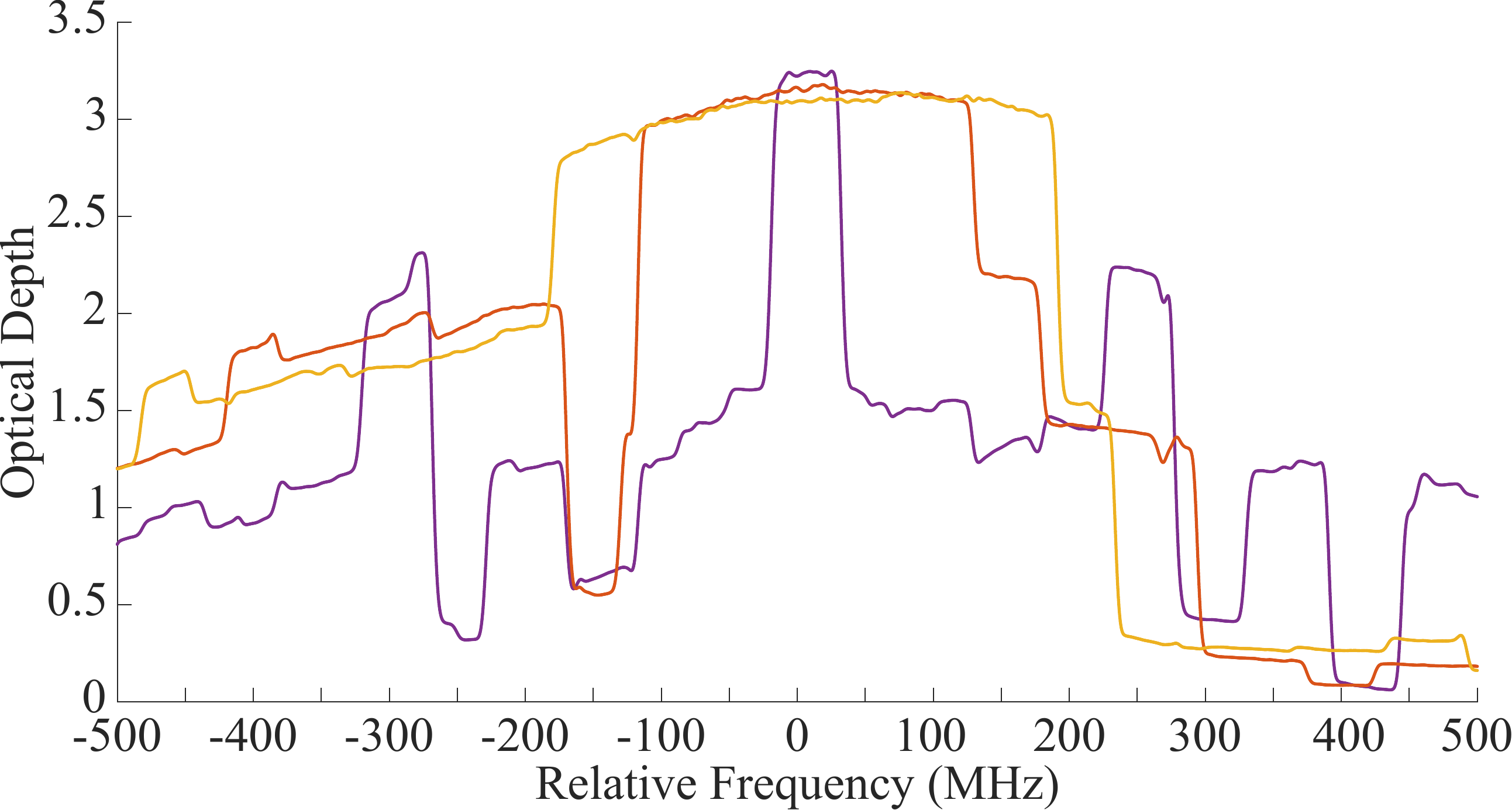}
    \caption{Close-up of the $\ket{4_g}-\ket{1_e}$ anti-hole, showing the optical depth after the SP at bandwidths of 50 MHz (purple), 250 MHz (red) and 360 MHz (yellow).}
    \label{fig:bdw}
\end{figure}

We also studied the SP for varying bandwidths, using otherwise the same parameters as above. Figure \ref{fig:bdw} shows the anti-hole on the $\ket{4_g}-\ket{1_e}$ transition at different bandwidths. The peak optical depth at 50 MHz is of 3.24, while at 360 MHz it is of 3.09. The small difference (less than 5 \%), for an increase in bandwidth by a factor of over 7, shows that the optical pumping is well saturated. We can also observe that the spectral features of the anti-hole remain flat. It should be noted, however, that while SP can be performed for bandwidths larger than 288 MHz, such bandwidths are not compatible with the CC required for AFC spin-wave storage. Nevertheless, our results show that SP can be performed efficiently for at least 288 MHz.

The optically pumped hyperfine states thermalize through spin relaxation processes. In \ref{sec:T1}, we show a measurement of the decay of the anti-hole at $\nu_{4_g,1_e}$, as well as the hole at $\nu_{1_g,1_e}$, as a function of the delay between the measurement and the end of the SP, for a bandwidth of 50 MHz. The decay fits well to a double-exponential function, with a fast $T_1^{\mathrm{fast}} = \SI{370 \pm 30}{\milli\second}$ and a slow $T_1^{\mathrm{slow}} = \SI{4.7\pm 0.2}{\second}$ decay constant. The anti-hole/hole lifetimes in the current 2 ppm crystal are significantly longer than those obtained in the 5 ppm crystal \cite{Businger2020}, indicating that the spin-spin flip-flop process had a strong influence on the hyperfine relaxation rates in that crystal. A more detailed study of the hole lifetimes will be presented in a separate study.

\subsection{Broadband and efficient AFC echoes} \label{sec:bb_eff_echoes}

\begin{figure}[t]
    \centering
    \includegraphics[width=1\linewidth]{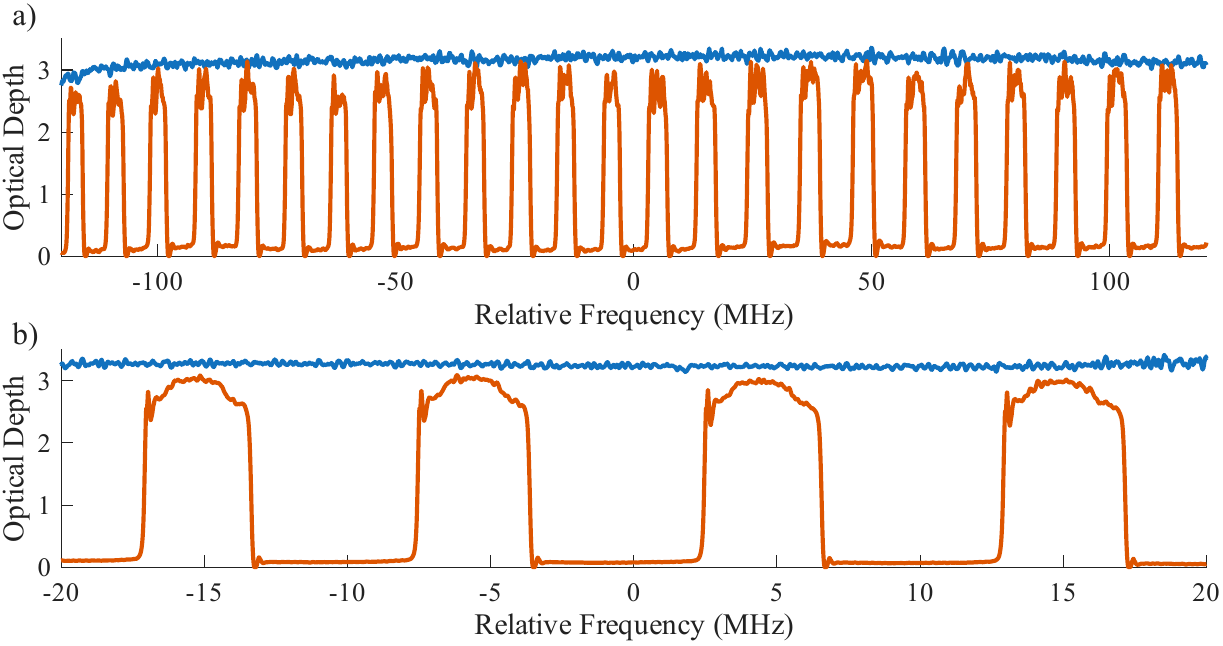}
    \caption{a) Measured optical depth of a typical AFC (red curve) burned on the $\ket{4_g}-\ket{1_e}$ anti-hole created through the SP, which can be compared to the measured anti-hole (blue curve) before the comb preparation, cf. Fig. \ref{fig:rep}(a) and Fig. \ref{fig:bdw}. The comb had a periodicity of $\Delta = \SI{10}{\mega\hertz}$, finesse $F=2.45$, and bandwidth $\Gamma_{\mathrm{AFC}} = \SI{250}{\mega\hertz}$. b) High-resolution scan of the same comb showing four teeth in detail.}
    \label{fig:teeth}
\end{figure}

After having optimized the SP, we study the creation of the AFC using the new method for generating the AFC preparation pulse discussed in Sec. \ref{sec:AFC_numerical_signal_tricks}. The AFC is created through a pulse train made of 16 repetitions of a 4 ms long AFC preparation pulse, which was kept constant for all the AFC experiments shown in this work. The AFC preparation pulse amplitude and the programmed AFC finesse are optimized for each comb, to maximize the resulting AFC echo efficiency.

In Fig. \ref{fig:teeth}(a), we show the resulting experimental comb for a programmed comb bandwidth of $\Gamma_{\mathrm{AFC}} = 250~\mathrm{MHz}$ and storage time of $1/\Delta = 100~\mathrm{ns}$. At this short storage time, the peak optical depth of the comb almost reaches the optical depth after the SP, only a small fraction of the peak optical depth is lost during the AFC preparation, due to unwanted burning on the AFC teeth.

From the experimental comb, we extract a peak optical depth of $d=2.75$, a comb finesse of $F=2.45$ and a background optical depth of $d_0 = 0.085$. The shape of the teeth is mostly squarish, as seen in Fig. \ref{fig:teeth}(b), with a slight distortion at the peak of the teeth. Overall, the short-duration storage comb is close to the desired optimum shape, demonstrating a precise control of the AFC preparation over a large bandwidth.
      
\begin{figure}[t]
    \centering
    \includegraphics[width=0.8\linewidth]{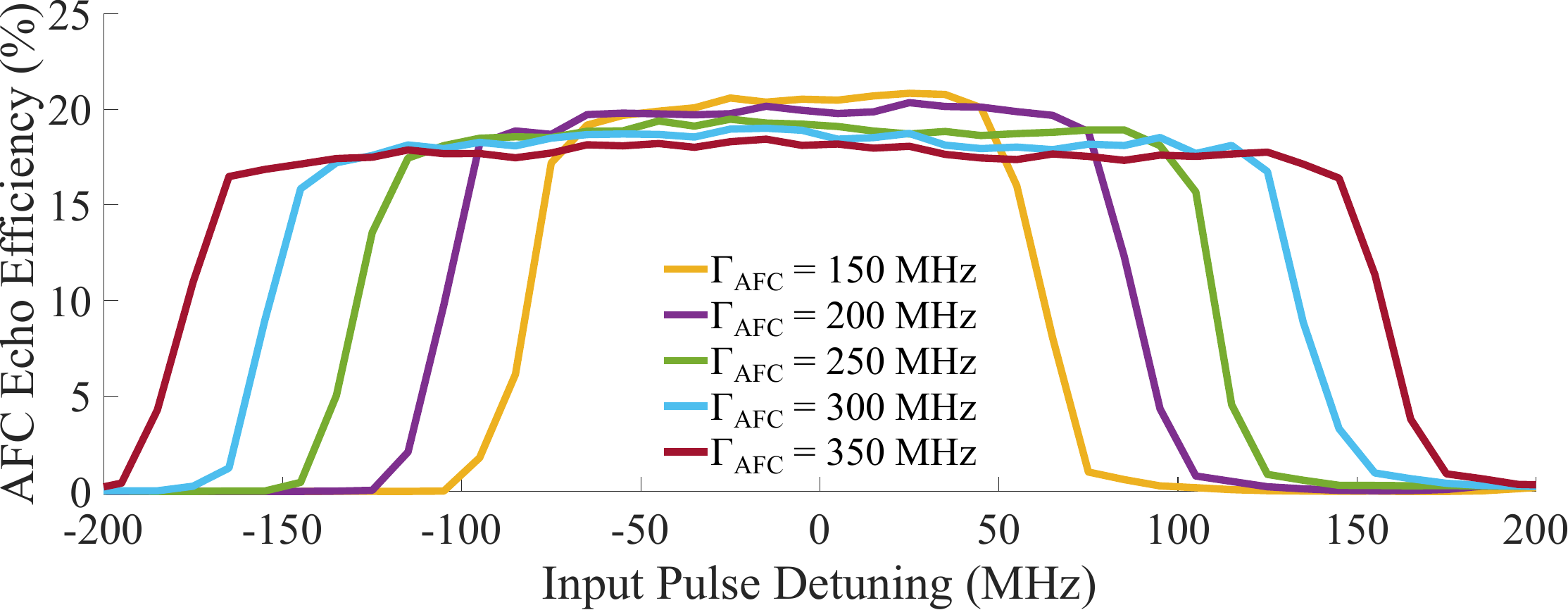}
    \caption{Measured AFC echo efficiency as a function of the detuning of the input pulse with respect to the center of the AFC, for AFC bandwidths ranging from $\Gamma_{\mathrm{AFC}} = 150~\mathrm{MHz}$ to $350~\mathrm{MHz}$. The storage time was set to $1/\Delta = 1~\mu\mathrm{s}$.}
    \label{fig:AFC_input_scan}
\end{figure}

We now turn to the measurement of the AFC echo efficiency. In Fig. \ref{fig:AFC_input_scan}, we show the measured echo efficiency at a short storage time of $1/\Delta = 1~\mu\mathrm{s}$, for different AFC bandwidths. The input pulse had a duration of $\sim40~\mathrm{ns}$, corresponding to a power spectrum bandwidth of $\sim 10~\mathrm{MHz}$. To study the efficiency across the entire AFC bandwidth, the carrier frequency of the input pulse was scanned over the AFC, recording the echo efficiency at each step of the scan, as shown in Fig. \ref{fig:AFC_input_scan}.

The measured echo efficiency is constant across the expected bandwidth, with little variation for the different programmed AFC bandwidths. For this short storage time, the echo efficiency at 150 MHz bandwidth is $20.5\%$, while it is $18.4\%$ at 350 MHz bandwidth. We can compare these values to the theoretical optimal AFC echo efficiency for the measured comb shown in Fig. \ref{fig:teeth}(a). Given its peak optical depth of $d=2.75$, the maximum attainable echo efficiency is $\eta_{\mathrm{AFC}} = 23.3\%$, see Eq. \ref{eq:afc_echo_efficiency}, for the optimal finesse of $F=2.71$. Taking into account the small background optical depth $d_0$, the theoretical echo efficiency drops to $\eta_{\mathrm{AFC}} e^{-d_0} = 21.4\%$ \cite{Riedmatten2008}, in excellent agreement with the measured efficiency. This indicates that our preparation method achieves close to optimal combs over these large bandwidths, similar to what has been achieved in, for instance, \euyso{} at much lower bandwidths of 5 MHz \cite{Ortu2022}.

\subsection{Long-lived AFC echoes}

The ability to generate long-lived AFC echoes, i.e. long $1/\Delta$ delays, is key for reaching a large temporal multimode capacity \cite{Afzelius2009a,Ortu2022}. Indeed, for a fixed bandwidth, i.e. a fixed input mode  duration, one can absorb many input modes, up until the echo of the first stored mode is produced after a delay $1/\Delta$.

\begin{figure}
    \centering
    \includegraphics[width=1\linewidth]{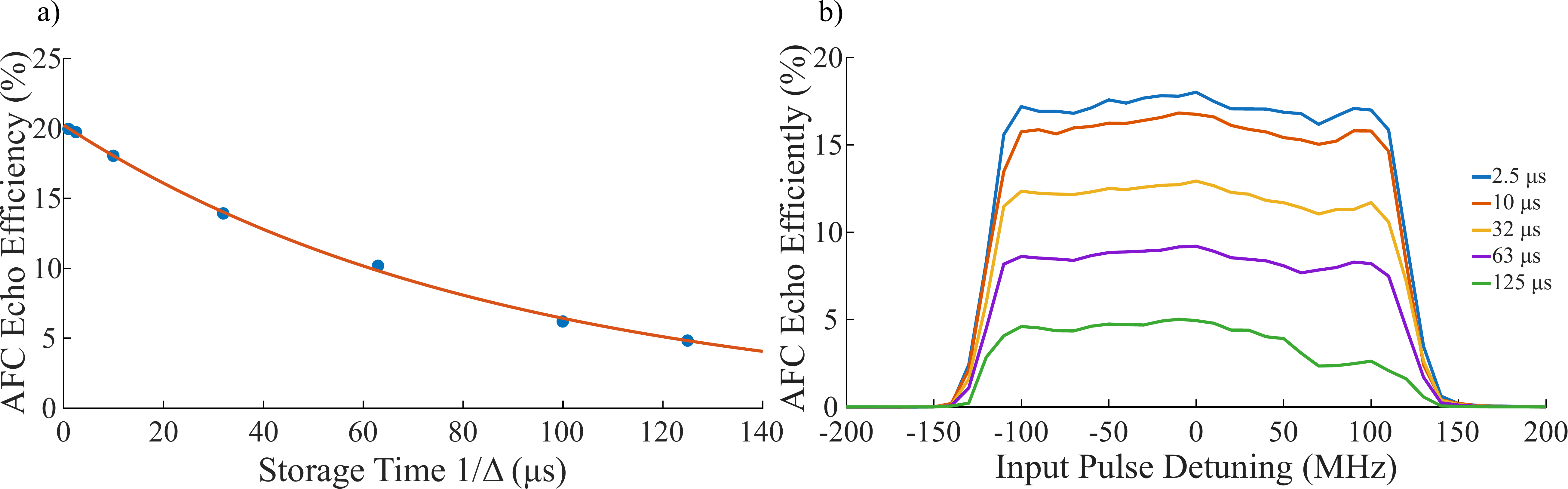}
    \caption{a) AFC echo efficiency as a function of the storage time, for a bandwidth of $\Gamma_{\mathrm{AFC}} = 250~\mathrm{MHz}$, measured at the centre of the AFC spectrum. The red line is the fit of the data, see text for details. b) AFC echo efficiency as a function of the detuning of the input pulse with respect to the centre of the AFC, for an AFC bandwidth of 250 MHz bandwidth. The measurement was repeated for $1/\Delta$ storage times ranging from $2.5~\mu\mathrm{s}$ to $125~\mu\mathrm{s}$.}
    \label{fig:AFC_T2}
\end{figure}

In Figure \ref{fig:AFC_T2}, we show the measured AFC echo efficiency as a function of the AFC delay $1/\Delta$, for a bandwidth of $\Gamma_{\mathrm{AFC}} = 250~\mathrm{MHz}$, showing the typical exponential decay. The decay time of the echo can be quantified by an effective AFC coherence time $T_2^{\mathrm{AFC}}$, through the decay function \cite{Jobez2016}

\begin{equation} \label{eq:AFC_T2}
    \eta_{\mathrm{AFC}} = \eta_{\mathrm{AFC}}(0) e^{-\frac{4}{T_2^{\mathrm{AFC}}\Delta}},
\end{equation}

\noindent where $\eta_{\mathrm{AFC}}(0)$ is the zero-delay efficiency. Fitting the function to the data in Fig. \ref{fig:AFC_T2}(a), we find an AFC coherence time of $T_2^{\mathrm{AFC}} = 348 \pm 15~\mu\mathrm{s}$ and a zero-delay efficiency of $\eta_{\mathrm{AFC}}(0) = 20.2~\%$. This is significantly longer than the AFC coherence time of $69 \pm 6$ $\mu$s we recently obtained in \cite{Businger2022}, using a \ybyso{} crystal with 5 ppm of doping concentration. The improvement is mostly due to technical changes made to the laser lock setup, which yields a narrower laser linewidth, and the longer optical coherence time in the 2 ppm sample \cite{Nicolas2023}, though we still not limited by the optical coherence time of $1~\mathrm{ms}$. The measured AFC coherence time compares very favourably to the ones obtained in non-Kramers ions \cite{Ortu2022}, where $T_2^{\mathrm{AFC}} = 92 \pm 6~\mu\mathrm{s}$ has been obtained in \pryso{}, and $T_2^{\mathrm{AFC}} = 249 \pm 14~\mu\mathrm{s}$ has been achieved in \euyso{}. It should be noted that the AFC memories in these non-Kramers ions have bandwidths below 10 MHz, which implies significantly lower temporal multimode capacity, for the same AFC delay.

In a final set of experiments, we study more in detail the AFC efficiency decay across the AFC bandwidth. In Fig. \ref{fig:AFC_T2}(b), we show the efficiency as a function of the input pulse detuning, for several $1/\Delta$ delays. The efficiency stays constant over the entire AFC bandwidth, for up to 125 $\mu$s. For the long 125 $\mu$s delay, the efficiency of the higher frequency components of the AFC decays slightly faster. We believe this is due to the higher storage times having much lower tolerances for imperfections in the AFC pulses, as is explained in \ref{sec:AFC_calibration}.

\section{Discussion and outlook}
\label{sec:discussion_outlook}

In this work, we have proposed and simulated an optical pumping scheme designed for long-duration AFC spin-wave storage experiments for up to 288~MHz bandwidth, with potential storage times longer than 1~ms \cite{Businger2020} in future experiments. The spin polarisation part of the scheme was explored experimentally, reaching 80\% of the hyperfine population into a single state. We also presented an efficient method for synthesizing broadband and high-resolution optical combs, allowing us to burn optimal AFCs for short storage times, reaching 20\% efficiency, and to achieve a storage time of 125~$\mu$s with 5\% efficiency. The corresponding AFC coherence time of 348~$\mu$s is longer than those achieved in non-Kramers ions, which so far have achieved the longest optical storage times based on AFC memories, while having a bandwidth that is at least 25 times larger. 

The near-term outlook is to use the model presented in Ref. \cite{Chen2025} to simulate the achievable AFC spin-wave bandwidth in \ybyso{}, taking into account relevant experimental parameters such as the Gaussian beam parameters of the input and control fields, the optical transition dipole moments, input and control pulse durations, optical control power and the memory bandwidth. These simulations will guide the experimental efforts to achieve millisecond storage time, for the highest bandwidth possible, thereby maximizing the temporal multimode capacity.

We have also presented a laser setup that is very frequency-agile and modular as it allows one to change the transitions accessed and the bandwidths used through programming of an AWG and MW sources. The simple optical setup using a single laser and a single, fibre-coupled EOM, and the use of off-the shelf, standard MW and RF components, offers the opportunity to create a rack-mounted version for future applications, an important aspect for developing a transportable quantum memory setup.

\section*{Acknowledgments}

This work has received funding from the Swiss State Secretariat for Education, Research and Innovation (SERI), under contract number UeM029-7.

\appendix
\setcounter{section}{0}

\section{Simulation of bandwidth limitation of CC and SP.}
\label{sec:BWlimit}

\begin{figure}[h]
    \centering
    \includegraphics[width=0.5\linewidth]{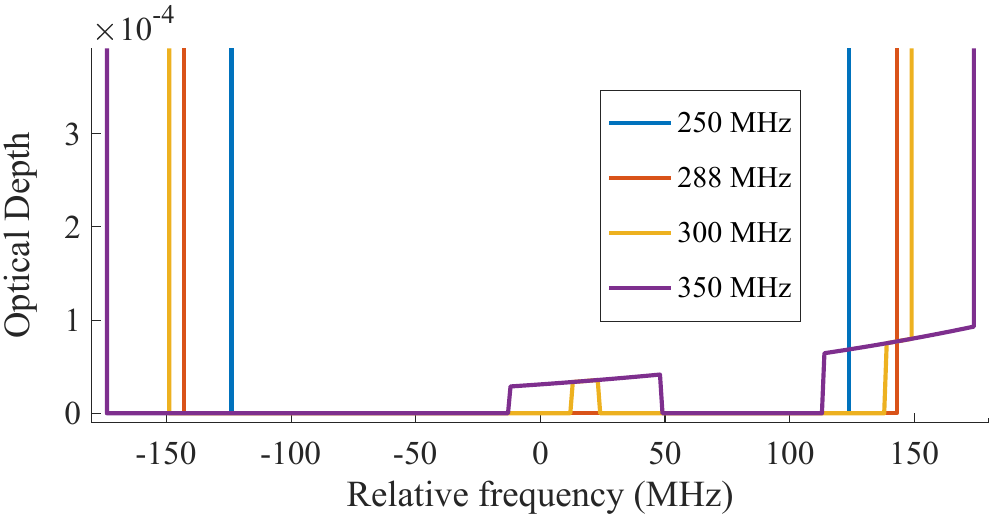}
    \caption{Difference between the optical depth of the absorption of the whole  ensemble and the one of the ions chosen for storage for different CC and SP bandwidths.}
    \label{fig:BWlimit}
\end{figure}

The figure \ref{fig:BWlimit} represents the difference between the spectrum including all frequency classes in the inhomogeneous ensemble (plotted in blue in Fig. \ref{fig:CC_scheme}(c)) and the spectrum due only to the selected frequency classes involved in the spin-wave storage (plotted in red in Fig. \ref{fig:CC_scheme}(c)), for different CC and SP bandwidths. We can see that, when the bandwidth is larger than 288 MHz, which corresponds to the splitting between the two hyperfine states $\ket{3_e}$ and $\ket{4_e}$, frequency classes not involved in the spin-wave storage start contributing to the absorption across the AFC bandwidth. This parasitic absorption would interfere with the spin-wave protocol and decrease its efficiency. It therefore sets an upper limit on the useful bandwidth for AFC spin-wave storage.

\section{Construction of the AFC burning pulse spectral phase} \label{sec:AFC_spectral_phase}

We proceed to derive the spectral phase Eq. \ref{eq:phase_new_method} of the AFC-burning pulse. In this appendix, all the frequency variables are given in $\SI{}{\radian\per\second}$, even those that reuse labels that were used for natural frequencies in the main text.

Let $s(t) = a(t) e^{i \phi(t)}$ be the analytic representation of the real-valued time-domain signal used to burn the AFC, and let its Fourier transform be denoted by $\hat{s}(\omega) = u(\omega) e^{i \theta(\omega)}$, with $a(t)$ and $u(\omega)$ the non-negative temporal and spectral envelopes, and $\phi(t)$ and $\theta(\omega)$ the temporal and spectral phases, respectively. We want to construct a signal $s(t)$ that is such that its spectral envelope is a rectangular frequency comb Eq. \ref{eq:ideal_AFC_burning_PSD}, and such that the peak amplitude $A$ of its temporal envelope $a(t)$ is minimized at constant average signal power $\bar{P} = \lim_{T_0 \rightarrow \infty} \frac{1}{2T_0}\int_{-T_0}^{T_0} a(t)^2 \dd{t}$. Equivalently, we wish to minimize the crest factor of the temporal signal, which is the ratio of the peak amplitude and the RMS power, $\text{CF} = A / \sqrt{\bar{P}}$. This problem of crest factor minimization is well known in the literature, and has been tackled both analytically \cite{Schroeder1970} and through iterative algorithms \cite{Yang_2015}.

To achieve this goal of crest factor minimization, we get to optimize $a(t)$, and either of the two phases $\phi(t)$ or $\theta(\omega)$ (once one is determined, the other is obtained via Fourier transformation). Ideally, we want $u(\omega)$ to be exactly the desired frequency comb, but the obtained $a(t)$ could just be an approximation of a desired optimal target temporal envelope. A natural choice of target envelope is a rectangular window of duration $T$, since this is the waveform with the minimum achievable $\text{CF}  = 1$. The goal is now to construct $\theta(\omega)$ such that $u(\omega)$ is exactly the desired spectral envelope, and $a(t)$ is approximately a rectangular envelope. This is essentially the problems of algorithmic (e.g. via the Gerchberg–Saxton\cite{Gerchberg1972APA} algorithm) and analytical phase retrieval \cite{Fowle1964}. In this work, we are interested in analytical solutions for the sake of fast numerical computation of the signals. Note that this problem is also similar to the Slepian concentration problem \cite{Slepian1961Bell}.

As in Ref. \cite{Fowle1964}, we approximate the inverse Fourier transform of $\hat{s}(\omega)$ using the stationary phase approximation \cite{erdelyi1956asymptotic},
\begin{align}
    s(t) & = \frac{1}{\sqrt{2\pi}} \int_{-\infty}^\infty u(\omega) e^{i \theta(\omega)} e^{i \omega t} \dd{\omega} \label{eq:iFT} \\ 
    & \approx \sum_{\omega_0 \in \Omega(t)} \frac{u(\omega_0)}{\sqrt{\abs{\theta''(\omega_0)}}} \exp \left[ i \left(\theta(\omega_0) + \omega_0 t + \frac{\pi}{4} \sgn \theta'' (\omega_0)\right) \right], \label{eq:SPA_iFT}
\end{align}
where $\Omega(t) = \set{\omega | \theta'(\omega) = -t}$ is the set of stationary points of the integrand of the Fourier integral Eq. \ref{eq:iFT}. We now restrict ourselves to solutions $\theta$ that are such that $\theta'$ is monotonic, so that there exists a unique stationary point $\omega_0(t)$ at any given time $t$. Equating the RHS of Eq. \ref{eq:SPA_iFT} to $s(t) = a(t) e^{i \phi(t)}$, we obtain the two approximate Fourier relations
\begin{align}
    a(t) & = \frac{u(\omega_0)}{\sqrt{\abs{\theta''(\omega_0)}}}\label{eq:SPA_Fourier_pair_envelope},\\
    \phi(t) & = \omega_0 t + \theta(\omega_0) + \frac{\pi}{4} \sgn \theta''(\omega_0).\label{eq:SPA_Fourier_pair_phase}
\end{align}
Note that in particular using Eq. \ref{eq:SPA_Fourier_pair_phase} we have $\phi'(t) = \omega_0' t + \omega_0 + \omega_0' \theta'(\omega_0) = \omega_0$, i.e. the instantaneous frequency of $s$ is given by the stationary point $\omega_0$. Thus, requiring that $\theta'$ be monotonic implies that the time-domain signal is a monotonic chirp. Now, since we've assumed that $\theta'$ is monotonic, we have that $\theta''$ is either non-negative or non-positive, so we can write
\begin{equation*}
    \abs{\theta''(\omega)} = \pm \theta''(\omega),
\end{equation*}
where the $\pm$ sign depends on whether $\theta'(\omega)$ is chosen to be monotonically increasing or decreasing. 

The full derivation for an arbitrary $a(t)$ is done in Ref. \cite{Fowle1964}. For the simple case of a rectangular envelope $a(t) = A \rect{t/T}$ that we are interested in, we obtain from Eq. \ref{eq:SPA_Fourier_pair_envelope} for $\abs{t} \leq T / 2$ the differential equation
\begin{equation*}
    \theta''(\omega) = \pm \frac{1}{A^2} u^2(\omega),
\end{equation*}
were we've replaced the dependence on $\omega_0(t)$ by a dependence on $\omega$ without loss of generality, since there's no longer an explicit dependence on $t$. Solving this differential equation, and dropping the terms that are constant and linear in $\omega$ since they only lead to irrelevant phase shifts and time delays respectively, we find
\begin{equation*}
    \theta(\omega) = \pm \frac{1}{A^2} \int^\omega \int^{\omega}u^2(\omega) \dd{\omega} \dd{\omega}.
\end{equation*}
Furthermore, we can replace the amplitude $A$ by using Plancharel's theorem
\begin{equation*}
    \int_{-\infty}^\infty a^2(t) \dd{t} = A^2 T = \int_{-\infty}^\infty u^2(\omega) \dd{\omega},
\end{equation*}
so that
\begin{equation}\label{eq:SPA_phase_rectT}
    \theta(\omega) = \pm \frac{T}{\int_{-\infty}^\infty u^2(\omega) \dd{\omega}} \int^\omega \int^{\omega} u^2(\omega) \dd{\omega} \dd{\omega}.
\end{equation}
The equivalent result for signals that are periodic in time is derived (using a different approach) in Ref. \cite{Schroeder1970} (Eq. (10)). This spectral phase ensures that the temporal envelope is approximately rectangular of duration $T$. This approach gives good results for as long as the time-bandwidth product of the signal remains large, as is argued in Ref. \cite{Fowle1964}. As an example, for a rectangular spectral envelope of bandwidth $B$ of amplitude $U$ centred on a carrier $\omega_c \geq B/2$, $u(\omega) = U \rect{((\omega - \omega_c) / B)}$, we find
\begin{equation}\label{eq:SPA_phase_rectT_rectF}
    \theta(\omega) = \pm \frac{T}{2 B} \omega^2.
\end{equation}
In particular, if the time-domain signal is periodic with period $P$, as is the case when working with discrete Fourier transforms, the frequency components take discrete values that are integer multiples of $2\pi/P$, i.e. $\omega_n = 2\pi n / P$ and $B = 2 \pi M / P$, so that
\begin{equation} \label{eq:SPA_phase_rectT_rectF_periodic}
    \theta_n = \pm \pi \frac{T}{P M} n^2.
\end{equation}
For the special case $T = P$, this is the same formula derived in Ref. \cite{Schroeder1970} (Eq. 12). In addition, to illustrate the point about the instantaneous frequency of the time-domain signal, note that using the stationary phase condition $\theta'(\omega_0) = - t$ and the result $\phi'(t) = \omega_0$, we find that $\phi'(t) = \pm \tfrac{B}{T} t$, i.e. $\theta(\omega)$ tries to turn the time-domain signal into a linear chirp. 

For a general $u(\omega)$, such as the desired frequency comb of bandwidth $\Gamma_\text{AFC} = N \Delta$, 
\begin{equation*} 
    u(\omega) \propto \sum_{n = 0}^{N-1} \rect \frac{\omega - \sigma/2 - n \Delta}{\sigma}.
\end{equation*}
one can similarly compute the adequate $\theta(\omega)$, either analytically, or numerically in constant-$N$ time. This spectral phase does a good job at reducing the crest factor of the AFC signal for as long as the number of teeth is small, with crest factors of $\sim 1.4$ until about $100$ teeth, and about $2.5$ afterwards.

To understand why the crest factor increases after a certain number of teeth, one can recognize that the $\theta$ we've constructed tries to accomplish two things : (1) prevent the constructive interference between the $N$ individual teeth's impulse responses, which would produce a nascent Dirac comb (or more accurately, a Dirichlet kernel) with spacing $2\pi/\Delta$ if added with zero phase, and (2) prevent the tooth's impulse response itself from concentrating into a large-amplitude $\sinc$ function which modulates the Dirac comb's amplitude. (2) is actually a rather weak constraint, considering the tooth bandwidth will decrease at constant comb finesse as $N$ increases, and thus the $\sinc$ will spread out temporally and decrease in amplitude. Yet, it is (2) which causes problem. To see how, go back to the assumption that $s$ is a monotonic chirp : this assumption forces $\theta(\omega)$ to be such that, in the time domain, each tooth's chirp of bandwidth $\sigma$ happens sequentially one at a time for a duration $T / N$ each. This is fine as long as the time-bandwidth product $\sigma (T / N)$ is large, but when this is not the case Eq. \ref{eq:SPA_iFT} becomes a poor approximation and thus the actual temporal envelope $a(t)$ will not match the desired rectangular envelope, therefore increasing the crest factor.

To improve on this approach, one should then somehow relax the constraint that $\theta'$ be monotone. To do so, we exploit the fact that our desired $u(\omega)$ can be written as the product of two functions -- an infinite comb of rectangular teeth of period $\Delta$, and a rectangular envelope of width $\Gamma_\text{AFC}$. Formally, let's write $\hat{s}$ as a slowly-varying function $\hat{G}$ multiplied by a periodic function which is a periodic summation of a narrowbanded function $\hat{g}$ with period $\Delta$,
\begin{equation*}
    \hat{s}(\omega) = \hat{G}(\omega) \sum_{n=-\infty}^\infty \hat{g}(\omega - n\Delta),
\end{equation*}
where both $\hat{G}$ and $\hat{g}$ are complex-valued. Since we've assumed that $\hat{g}$ is narrow and $\hat{G}$ is slowly varying, we can write
\begin{equation*}
    \hat{s}(\omega) \approx  \sum_{n=-\infty}^\infty \hat{G}(n\Delta)\hat{g}(\omega - n\Delta),
\end{equation*}
i.e. the periodic function essentially samples $\hat{G}$, so that in the time domain we have
\begin{equation*}
    s(t) \approx g(t) \sum_{n=-\infty}^\infty \hat{G}(n\Delta)e^{i n\Delta t}.
\end{equation*}
For instance, for the case of interest where both $\hat{G}$ and $\hat{g}$ are rectangular, one can then construct $\hat{g}$'s spectral phase using Eq. \ref{eq:SPA_phase_rectT_rectF} such that $g$ is rectangular of duration $T$, and $\hat{G}$'s spectral phase using Eq. \ref{eq:SPA_phase_rectT_rectF_periodic} such that one period $2 \pi / \Delta$ of the Fourier series $\sum_{n=-\infty}^\infty \hat{G}(n\Delta)e^{i n\Delta t}$ is a rectangular envelope of duration $2\pi/\Delta$. Explicitly, for a tooth shape
\begin{equation*}
    \abs{\hat{g}(\omega)} = \rect \omega / \sigma,
\end{equation*}
one assigns the phase Eq. \ref{eq:SPA_phase_rectT_rectF} with $B = \sigma$ and $T$ the total duration of the rectangular pulse,
\begin{equation*}
    \arg \hat{g} = \frac{T}{2\sigma}\omega^2,
\end{equation*}
and for a slowly-varying envelope shape
\begin{equation*}
    \abs{\hat{G}(\omega)} = \rect \omega / \Gamma_\text{AFC},
\end{equation*}
one assigns to $\hat{G}_n \equiv \hat{G}(n\Delta)$ a phase $\theta_n$ that's constant over one $\Delta$ using Eq. \ref{eq:SPA_phase_rectT_rectF_periodic} for a bandwidth $\Gamma_\text{AFC} = N \Delta$ and a rectangular envelope of duration $2\pi/\Delta$, i.e. $P = T = 2\pi/\Delta$ and $M = N$ in Eq. \ref{eq:SPA_phase_rectT_rectF_periodic}, such that
\begin{equation*}
    \arg \hat{G}_n = \pi \frac{1}{N} n^2.
\end{equation*}
Putting everything together, one retrieves Eq. \ref{eq:phase_new_method}. With this construction, one essentially delegates the task of obtaining a long envelope of duration $T$ to the narrowbanded components, and the task of obtaining a short, periodic envelope of duration $2\pi/\Delta$ to the broadband component. The per-tooth term $\arg \hat{g}$ will only have a significant contribution to the overall temporal envelope when $\sigma$ is large compared to $1/T$. It will again do a poor job at making $g(t)$ have a rectangular envelope when the time-bandwidth product $\sigma T$ becomes small, but this does not matter because when $\sigma$ becomes small the $\sinc$ spreads out in time and its amplitude becomes small. On the other hand, the $\hat{G}$ phase term will do a good job at producing a rectangular envelope regardless of how $\hat{g}$'s phase performs, as long as there are many teeth, i.e. as long as the time-bandwidth product $N / (2\pi) = \Gamma_\text{AFC} / \Delta$ is large. Overall, this choice of phase tends to perform more poorly than the one constructed using Eq. \ref{eq:SPA_phase_rectT} for small numbers of teeth ($\text{CF} \approx 1.8$), but does better for many teeth ($\text{CF} \approx 1.8$).

One could then choose the appropriate construction for $\theta$ depending on which regime one's AFC-burning signal is in, Eq. \ref{eq:SPA_phase_rectT} for $N < 100$ or Eq. \ref{eq:phase_new_method} for $N > 100$. If one requires even lower crest factors, one could use further iterative crest factor reduction algorithms such as the one in Ref. \cite{Yang_2015} to lower it even more, using the $\theta$'s above as an initial guess for these algorithms. In practice, we found that doing this can decrease the crest factor to about $1.75$, but the significant computational overhead of these methods did not prove to be worth it.

In addition, if one needs to tune the shape of $\abs{\hat{G}}$, e.g. as a precompensation to the bandwidths of the waveform generator or subsequent amplifiers to improve power flatness across the entire band, one can always compute the proper phase $\arg\hat{G}$ using Eq. \ref{eq:SPA_phase_rectT} for the given $\abs{\hat{G}}$, and the obtained crest factor should not increase significantly as long as $\abs{\hat{G}}$ varies slowly over a period $\Delta$.

As a final note, one can always recover the time-domain phase from the above construction, and compute the signal in the time domain with a perfect rectangular envelope. One obtains the total temporal phase 
\begin{equation*}
    \arg{s(t)} = \frac{\sigma}{2 T} t^2 + \frac{\Gamma_\text{AFC}}{4\pi/\Delta} \left(\left(t \% \frac{2\pi}{\Delta}\right) - \frac{\pi}{\Delta}\right)^2.
\end{equation*}
This way, the crest factor will be the minimum of $1$, at the cost of distortion of the power spectral density -- in practice in the form of ripples both on the edges of the $\Gamma_\text{AFC}$ band and on the teeth themselves. In our case, this was not desirable, considering the AFC efficiency at long storage times $2\pi / \Delta$ is quite sensitive to power fluctuations, see Fig. \ref{fig:AFC_eff_optimization} (b).

\section{Spectral Anti-hole Lifetime}\label{sec:T1}

\begin{figure}[h]
    \centering
    \includegraphics[width=0.8\linewidth]{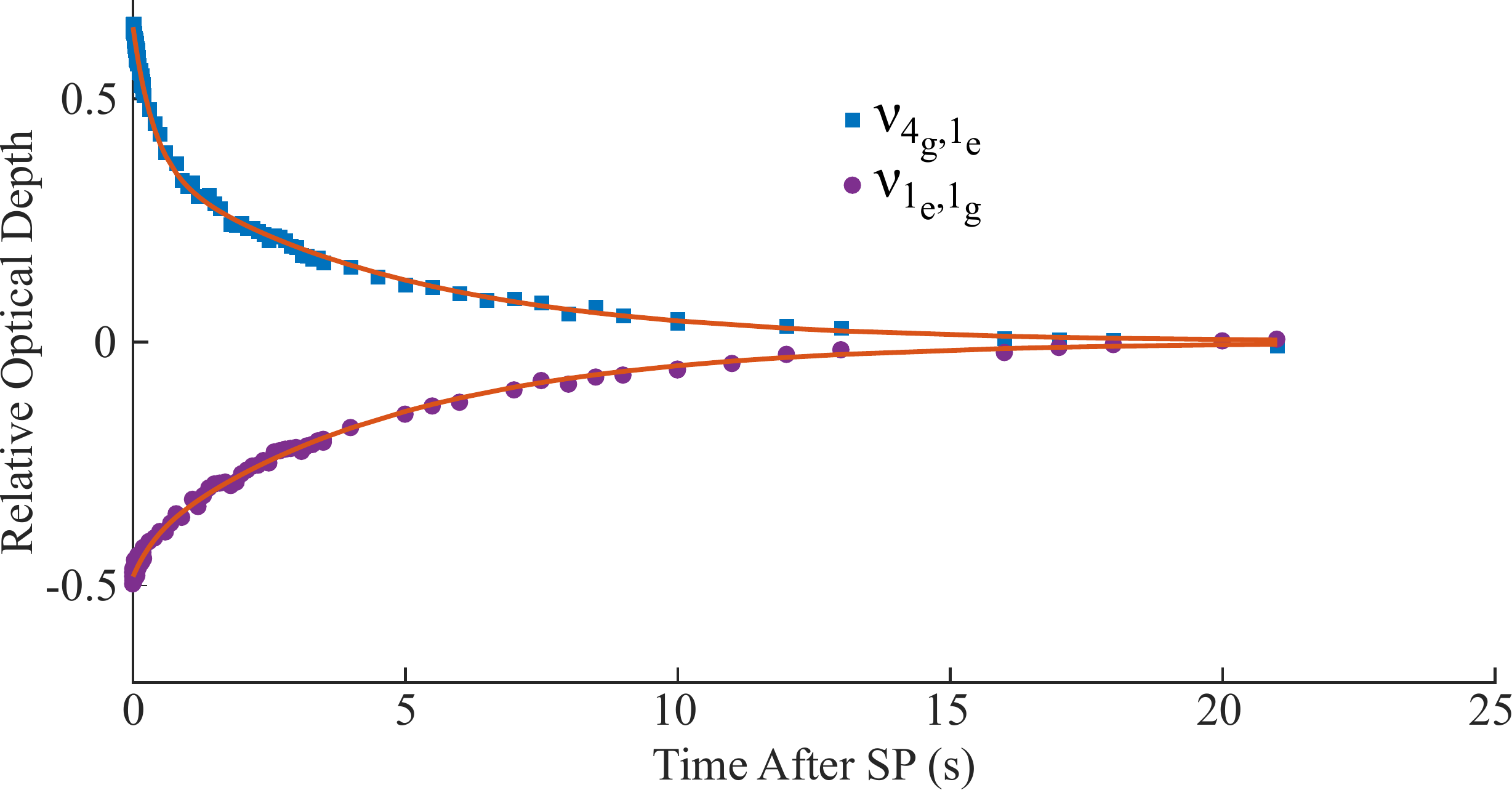}
    \caption{Data points: Optical depth, relative to the unpolarised absorption spectrum, measured at the frequencies $\nu_{4_g,1_e}$ and $\nu_{1_g,1_e}$ as a function of the delay between the SP and the scan of the absorption spectrum. Curves: double-exponential fit.}
    \label{fig:AFC_T1}
\end{figure}

Figure \ref{fig:AFC_T1} shows the decay of the spin population of the $\nu_{4_g,1_e}$ anti-hole and of the $\nu_{1_g,1_e}$ hole after a SP time of 300 ms and an anti-hole with a bandwidth of $50~\mathrm{MHz}$ when the \ybyso{} crystal is cooled to a temperature of $3.1~\mathrm{K}$. Each point is taken with a new SP sequence, then waiting a time $t$ to take a scan of the optical depth over the entire absorption spectrum as in \ref{fig:AFC_eff_optimization}(a). We subtract the optical depth from the unpolarised absorption spectrum from the optical depth measured, giving us a relative optical depth $d_{rel}(t)$. When the data is fit with a double exponential of the form
\begin{equation}
    d_{rel}(t) = Ae^{\frac{-t}{T_1^{\mathrm{fast}}}} + Be^{\frac{-t}{T_1^{\mathrm{slow}}}},
    \label{eq:AFC_T1}
\end{equation}
\noindent we get $T_1^{\mathrm{fast}} = 370 \pm 30~\mathrm{ms}$ and a slow $T_1^{\mathrm{slow}} = 4.7 \pm 0.2~\mathrm{s}$. These lifetimes are the main limitation of the efficiency of the SP. Improving these lifetimes would require further lowering the doping concentrations, at the expense of optical depth and AFC echo efficiency, or working at lower temperatures.

\section{AFC calibration curves}\label{sec:AFC_calibration}

\begin{figure}[h]
    \centering
    \includegraphics[width=1\linewidth]{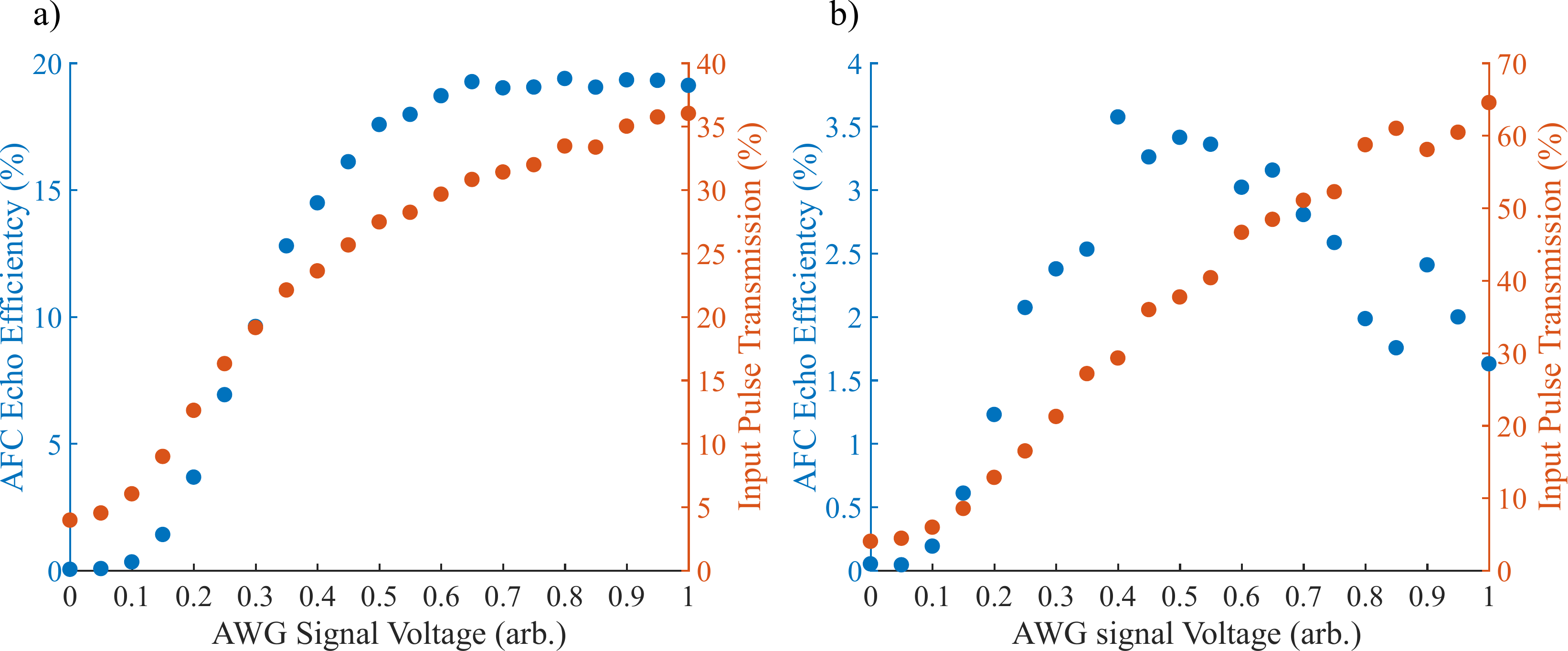}
    \caption{a) AFC echo RF calibration curve for a storage time of 1 $\mu$s b) AFC echo RF calibration curve for a storage time of 125 $\mu$s.}
    \label{fig:AFC_eff_optimization}
\end{figure}

The method used to calibrate the 250 MHz AFC memory can be seen in Fig. \ref{fig:AFC_eff_optimization} where the amplitude of the pulses created by our AWG is scanned to find the maximum AFC efficiency. The optical power used for the AFC is the same as used for the SP, at $\SI{16}{\milli\watt}$, but the amplitude of the EOM modulation of the light is modified in our calibration sequence. We probe the centre of the AFC band with classical pulses for these calibration measurements. At low storage times, as in Fig. \ref{fig:AFC_eff_optimization}(a), the efficiency of the AFC echo saturates and remains high over a large range of RF pulse power. This leads to AFC echoes that are efficient over their entire bandwidth and are resistant to variations in the pulse power both over time and the full bandwidth. At longer storage times, however, the efficiency forms a narrow peak as shown in Fig. \ref{fig:AFC_eff_optimization}(b). Despite how constant the spectral power density is over the entire bandwidth of the AFC pulses, different spectral areas of the pulse are attenuated differently through our system. This means that, while the centre of the AFC is calibrated correctly, the edges might receive too much or too little power. The former causes power broadening, decreasing the available optical depth for the echo, while the latter doesn't allow for the AFC teeth to be carved out efficiently.

\pagebreak

\bibliography{qmcommon}
\bibliographystyle{iopart-num}

\end{document}